\documentclass[preprint,aps]{revtex4}

\usepackage{graphicx}
\usepackage{dcolumn}
\usepackage{bm}

\begin{document}

\topmargin=0.1in

\preprint{}

\title{On the Penetrating Nature of Electromagnetic Probes}

\author{Kevin L. Haglin}
\homepage{http://feynman.stcloudstate.edu/haglin}
\affiliation{Department of Physics, Astronomy and Engineering
Science\\St. Cloud State University, St. Cloud, MN 56301, USA}

\date{\today}

\begin{abstract}
We establish the extent to which photons and dileptons are penetrating
probes of hot hadronic matter by computing mean free paths.  Absorption 
is indeed negligible for real photons
as mean free paths are roughly tenths of {\AA}ngstr\"oms.
However, vector dominance effects, which provide rich
structure in virtual photon mass spectra, introduce 
nonnegligible reasborption and rehadronization resulting in 
mean free paths of the order tens of femtometers.  Consequences 
are illustrated by considering the {\it net\/} dilepton yield near
the rho peak via thermal $\pi\pi$ annihilation from a short-lived
hadronic fireball.
Results suggest that disappearance of rho in heavy-ion collisions can 
be due in part to reabsorption/rehadronization rather than solely due to 
collective effects.
\end{abstract}


\maketitle

Ever since the suggestion more than twenty years 
ago\cite{ef76,es78,es80} that photons and dileptons would make excellent 
probes of hadronic matter owing to their expected long mean free paths, there 
has grown a wide body
of literature both on the theoretical side\cite{theory} assessing various
features of production rates and model calculations for yields, and on the
experimental side documenting a variety of results in heavy-ion collisions
spanning a wide range of energies and systems\cite{expt}.  
Electromagnetic signals
have small cross sections of the order $\alpha$ for
real photons ($\alpha^{2}$ for dileptons) times
the square of some characteristic hadron size, {\em i.e.\/} 1 fm$^{2}$, and are
therefore rarely produced as compared to hadronic signals with
cross sections of order 1--10 times the same characteristic
1 fm$^{2}$.  Once produced, they reinteract with the same small
probabilities and consequently travel through and indeed escape the 
nuclear matter carrying with them valuable kinematical fingerprints of 
the interior of the 
fireball.  Electromagnetic probes are today considered one of the premier
assessment tools for studying possible phase crossover from hadronic
matter to prehadronic matter, the so-called quark-gluon plasma.  Eventual 
success of
this endeavor is contingent in part on the purity of 
electromagnetic probes.  Up to now, only back-of-the-envelope
estimates have been made, so we undertake a more quantitative
study with particular emphasis on virtual photons with masses near 
vector mesons.

We begin by discussing the general features of thermal photon 
production mechanisms in hadronic matter and later consider
dileptons.  Real photons in the
energy range 500 MeV to a few GeV are produced in thermalized
matter via hadronic reactions, most notably, $\pi+\rho\rightarrow\pi+\gamma$ 
supporting either pion exchange or a resonant {\it a\/}$_{1}$
contribution\cite{joepeter,xiong}.   Bremsstrahlung mechanisms dominate 
the energy region
below a few hundred MeV.  Given that the photons are on shell, they
will neither hadronize nor decay electromagnetically and we need only focus 
on absorption.  
Their mean free path, assuming dominance from the reaction
$\gamma+\pi\rightarrow a_{1} \rightarrow \rho+\pi$, is 
\begin{eqnarray}
\langle\lambda\rangle & = & {1\over\overline{\Gamma}},
\end{eqnarray}
where
\begin{eqnarray}
\overline{\Gamma} & = & {1\over 8\pi^{2}T}
\int_{z_{0}}^{\infty}\,dz\,\lambda(s,0,m_{\pi}^{2})\,
{\cal K\/}_{1}(z)\,\sigma(s)
\end{eqnarray}
and
\begin{eqnarray}
\sigma(s) & = & {3\pi\over\,k^{2}}{m_{a_{1}}^{2}\Gamma_{a_{1}}^{2}\over\left(
s-m_{a_{1}}^{2}\right)^{2}+\left(m_{a_{1}}\Gamma_{a_{1}}\right)^{2}}
\left({4\pi\alpha\over\,f_{\rho}^{2}}\right),
\label{realxsect}
\end{eqnarray}
$k$ is the $\pi\rho$ center-of-mass momentum, $z$ = $\sqrt{s}/T$, $z_{0}$ =
$(m_{\pi}+m_{\rho})/T$, $\lambda(x,y,z)$ =
$x^2-2x(y+z)+(y-z)^{2}$---a function which accounts for relative velocity
and other kinematical effects, $f_{\rho}$ is the
``universal'' vector-dominance coupling constant for rho $\simeq$ 
5.1\cite{bhaduri}, 
and ${\cal K\/}_{1}$ is a modified
Bessel function.  Branching ratios and spin degeneracies have
been included to give the particular prefactor appearing 
in Eq.~(\ref{realxsect}).
An
energy-dependent $a_{1}$ width is used having 400 MeV at
the centroid\cite{xiong}, which gives the following results
for the mean free paths: 1.80, 0.08, and 0.02 {\AA}
at 100, 150 and 200 MeV temperature
respectively.
Other contributions can be introduced; for example, {\it t\/}-channel 
reactions involving
{\it a\/}$_{1}$, $\pi$-exchange and $\omega$-exchange graphs are
next-leading candidates,
but the inevitable result would be a smaller $\langle\lambda\rangle$ 
by at most a factor of 2--3 since the 
cross sections
for these additional processes are somewhat smaller than the one considered 
here.
With mean free paths several orders of magnitude larger than any nuclear
system, we safely conclude that real photons penetrate completely the
entire spacetime extent of the nuclear reaction.

\begin{figure}
\includegraphics{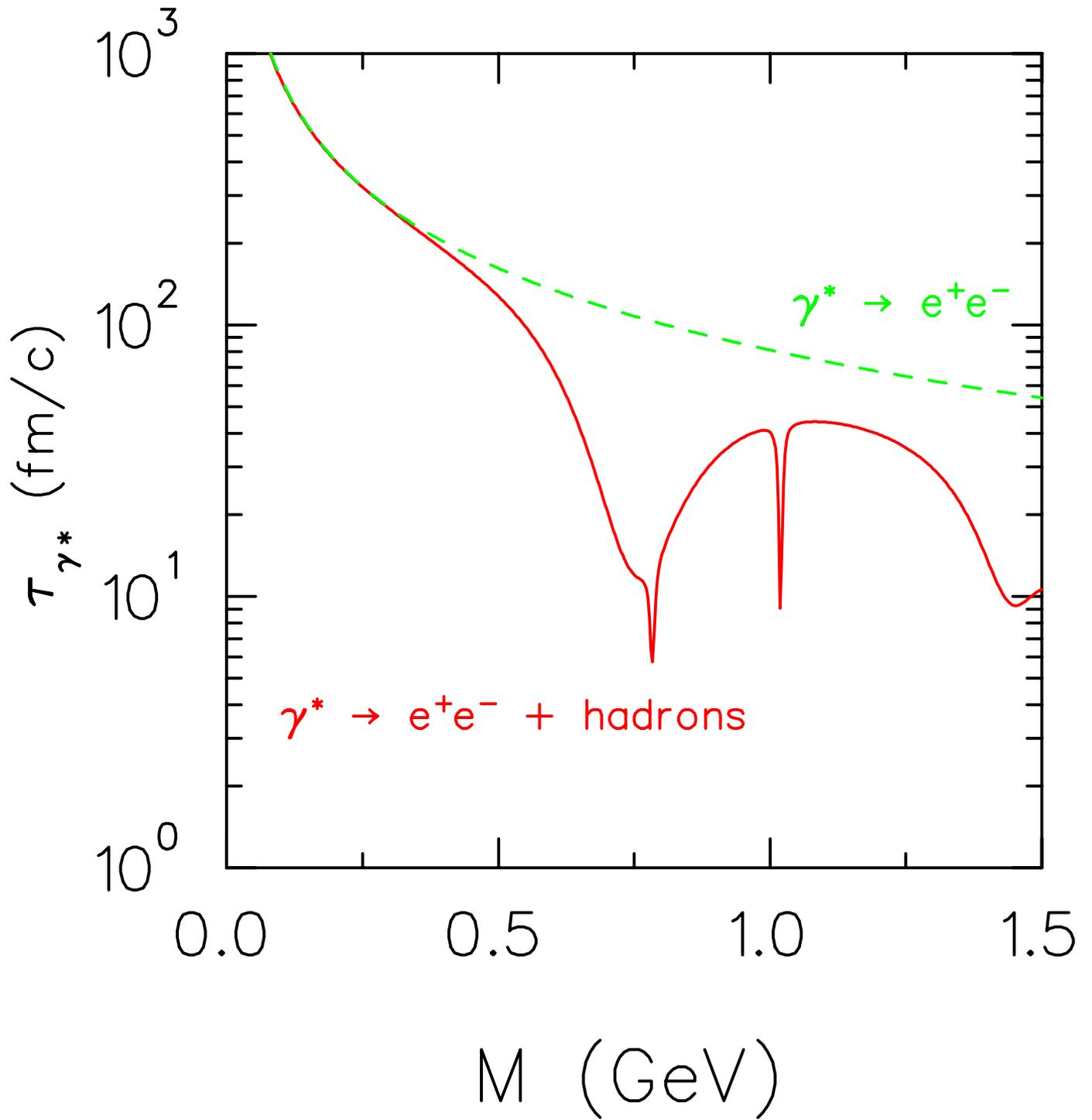}
\caption{\label{vtau} Virtual photon lifetime as a function of invariant mass.
             Dashed curve includes electromagnetic decays only, while
             solid curve includes in addition, contributions from $\rho$,
             $\omega$, $\phi$, and one of the $\rho^{\prime}$ states.}
\end{figure}

Photons off their mass shell (virtual photons) are thermally 
produced most abundantly
by $\pi\pi$ annihilation for masses a few hundred MeV to about 
1 GeV\cite{charlesjoe} and
via bremsstrahlung ($\pi\pi\rightarrow\pi\pi\gamma^{*}$)
as well as other hadronic reactions ({\it e.g.} 
$\pi\rho\rightarrow\pi\gamma^{*}$) for
invariant mass below about 300 MeV\cite{satz,khcgve,kh96}.
Virtual photons eventually decay
into lepton pairs or possibly back into hadrons. 
We consider a virtual photon
of mass {\it M\/} produced in the interior of the fireball.
Where does it decay?  The answer clearly depends on its own lifetime
against electromagnetic and/or hadronic decays.  The rate for decay
comes from two pieces: a purely electromagnetic rate plus a hadronic
decay (reverse vector dominance) as follows
\begin{widetext}
\begin{eqnarray}
\Gamma_{\gamma^{*}} & = & \Theta(M-2m_{\ell}){\alpha\,M\over 
3\/}\sqrt{1-{4m_{\ell}^{2}\over\,M^{2}}}
\left(1+{2m_{\ell}^{2}\over\,M^{2}}\right)
\nonumber\\
& \ & + \Theta(M-2m_{\pi})\left({4\pi\alpha\over f_{\rho}^{2}}\right)
{m_{\rho}^{4}\over(M^{2}-m_{\rho}^{2})^{2}+(m_{\rho}\Gamma_{\rho})^{2}}
\,\Gamma_{\rho},
\label{gswidth}
\end{eqnarray}
\end{widetext}
and we use an energy-dependent $\rho\rightarrow\pi\pi$ width
\begin{eqnarray}
\Gamma_{\rho} & = & {g_{\rho\pi\pi}^{2}\over\,4\pi}{2\over\,3}
{p^{3}\over\,m_{\rho}^{2}},
\end{eqnarray}  
with $g_{\rho\pi\pi} = 6.0$ and $p$ being the center-of-mass
momentum of the decay products.
Other resonances can
also be included in Eq.~(\ref{gswidth}), {\it e.g.} $\omega$ and $\phi$ open up
at the three-pion threshold, $\rho^{\prime}$ at the four-pion threshold, and
so on.
In Fig.~\ref{vtau}, we show the inverse width (lifetime) for
the virtual photon.
We conclude that once produced, the virtual photon carries 
kinematical information tens of femtometers at least before decaying.
It therefore decays, on average, 
{\it outside\/} the fireball! This does not mean that in-medium 
hadron properties are lost in favor of vacuum properties (unless, of
course, it re-hadronizes outside),
it merely sets the scale for decay.

\begin{figure}
\includegraphics{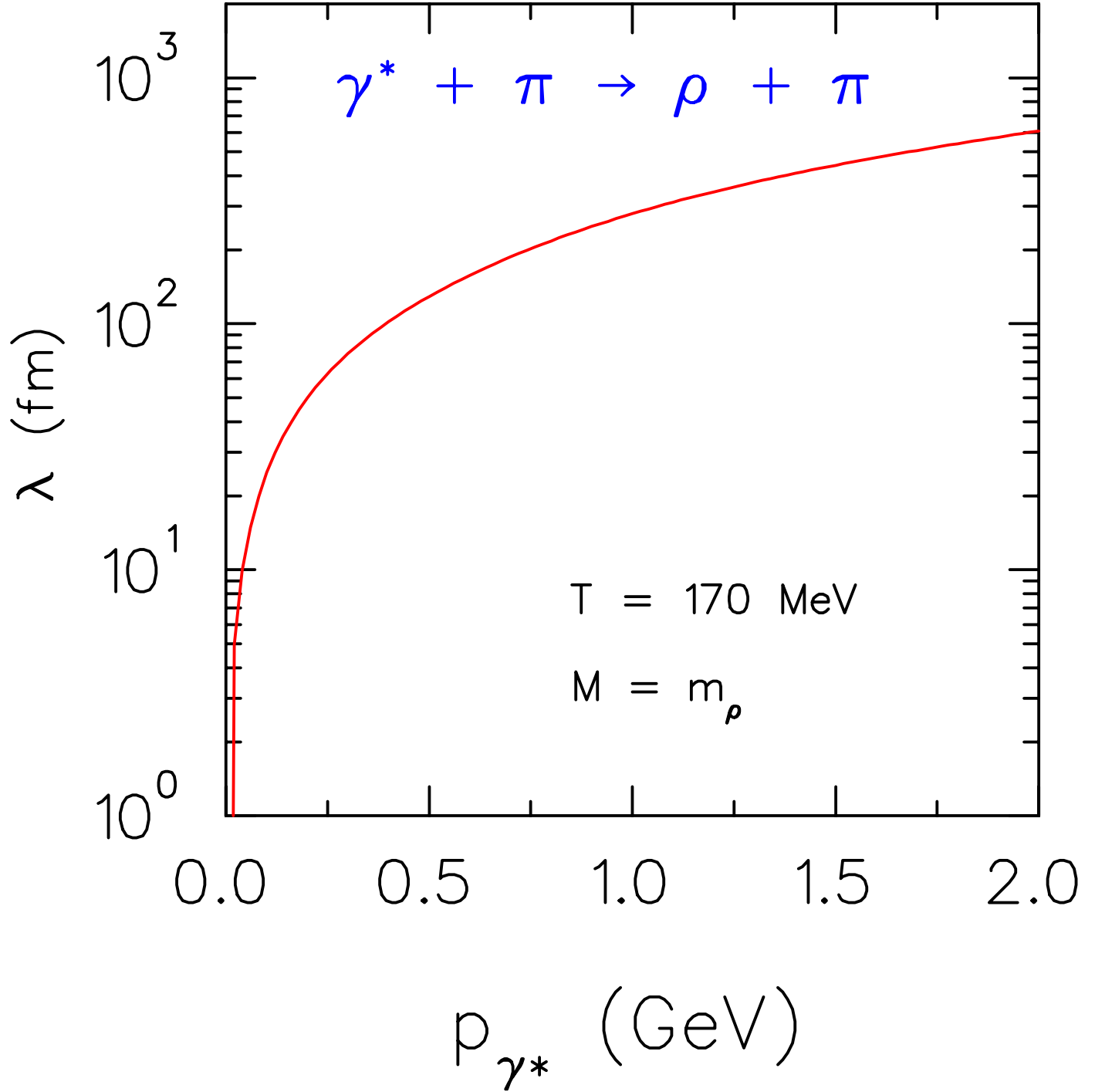}
\caption{\label{mfp} Free path for $\gamma^{*}$ having mass $m_{\rho}$
             at fixed temperature {\it T\/} = 170 MeV.}
\end{figure}

Since virtual photon propagation persists during the entire fireball
evolution and beyond, we must ask about the
chances that along the way it gets reabsorbed or possibly rehadronizes.
The free path as a function of temperature, invariant mass and
momentum is the following
\begin{eqnarray}
{\lambda} & = & {{v\/}\over\Gamma\/},
\end{eqnarray}
where  {\it v\/} = $p/E$, and the differential absorption rate $\Gamma$ is
\begin{eqnarray}
d\,\Gamma & = & g_{\pi}\,{d^{3}\/p_{\pi}\over(2\pi)^{3}}
f(E_{\pi})\,\sigma(s)\,v_{rel}
\delta\left((p+p_{\pi})^{2}-s\right)\,ds,\quad
\label{dgamma}
\end{eqnarray}
where
\begin{eqnarray}
v_{rel\/} & = & {\sqrt{(p\cdot\,p_{\pi})^{2}-M^{2}\,m_{\pi}^{2}}
\over\,E\,E_{\pi}},
\end{eqnarray}
$g_{\pi}$ = 2 reflecting the two possible charged states for the
entrance-channel pseudoscalar in $\gamma^{*}\pi\rightarrow\rho\pi$, $f$ is 
taken to be a Boltzmann and, where again, 
dominance of the $s$-channel $a_{1}$-exchange process
is assumed.  The
cross section is therefore simply obtained by 
multiplying Eq.~(\ref{realxsect}) by a vector dominance model (VDM) form 
factor
\begin{eqnarray}
|F_{\rho}(M)|^{2} & = & {m_{\rho}^{4}\over\left(M^{2}-m_{\rho}^{2}\right)^{2}
+(m_{\rho}\Gamma_{\rho})^{2}},
\end{eqnarray}
and by 2/3 to reflect the fact that a virtual photon has three
polarization states as compared with two for an on-shell photon.
From Fig.~\ref{vtau} it is clear that maximal effect occurs
near the vector poles.  So, in Fig~\ref{mfp}
we carry out the necessary integration in Eq.~(\ref{dgamma}) and plot 
the free path for a 
virtual photon at the rho mass at fixed temperature 
$T$ = 170 MeV as a function of its momentum.
Low momentum $\gamma^{*}$'s have reduced survival probabilities
due to the moderate mean free paths ranging from 1 fm to a few tens of 
femtometers.
In experiment, where cuts on dilepton three-momenta are made and
bins with a few hundred MeV momentum are analyzed, such reabsorption
could play an important role.

\begin{figure}
\includegraphics{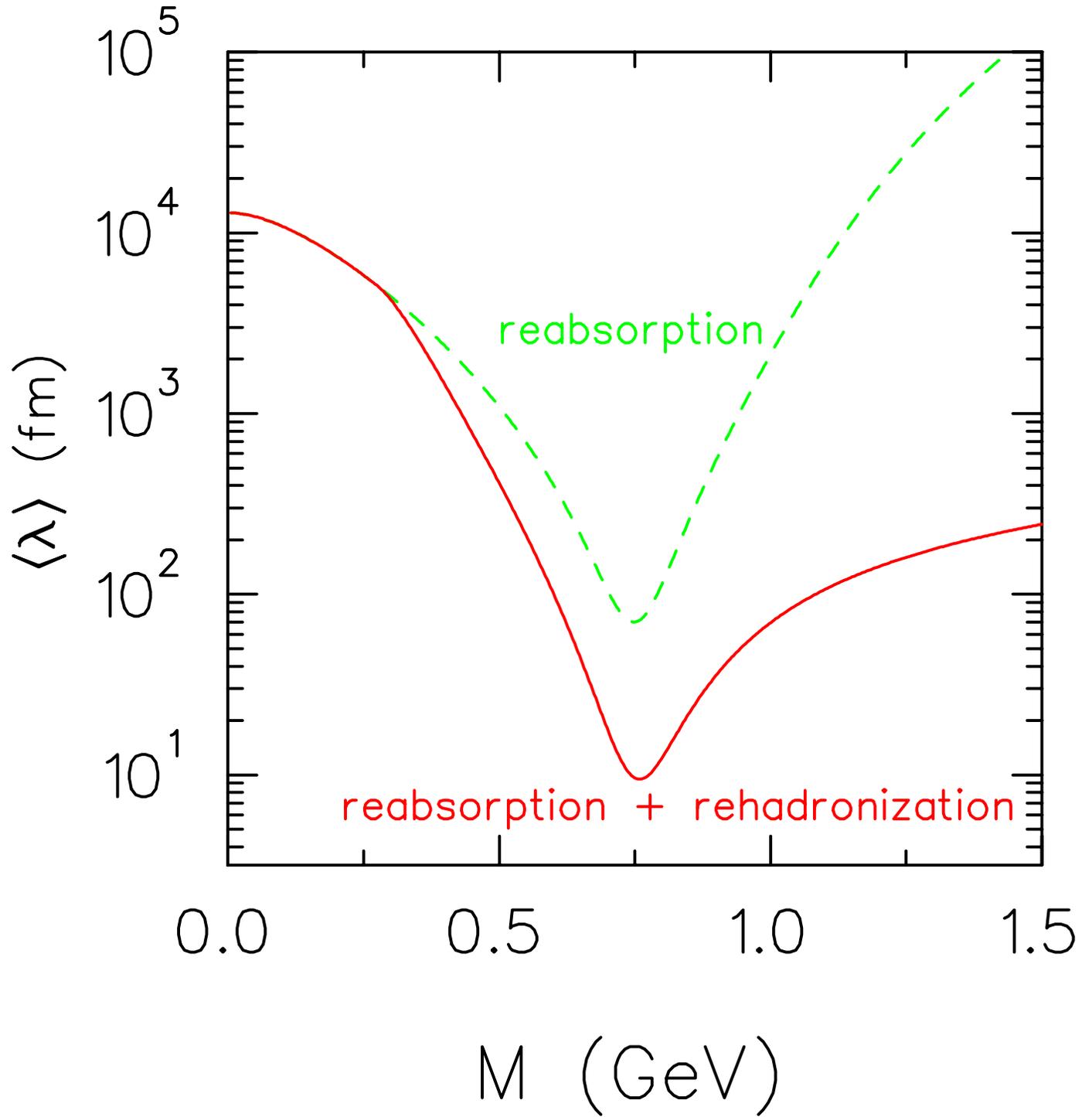}
\caption{\label{virtualmfp} Mean free path for $\gamma^{*}$ as a function of 
             mass at fixed tempertuare {\it T\/} = 170 MeV.}
\end{figure}

Next we average over virtual photon momentum (assumed to be distributed
according to a Boltzmann) and arrive at a mean free
path at fixed temperature as a function of mass alone.  The expression
can be written as
\begin{eqnarray}
\langle\lambda\rangle & = & {2\,e^{-M/T}\,T\left(M+T\right)\over
M^{2}{\cal K\/}_{2}(M/T)}{1\over\,\overline{\Gamma}^{\,abs}},
\end{eqnarray}
where ${\cal K\/}_{2}$ is again a modified Bessel function
and the absorption rate is
\begin{eqnarray}
\overline{\Gamma}^{\,abs} & = & {g_{\pi}T\over 
8\pi^{2}\,M^{2}{\cal K\/}_{2}(M/T)}
\int_{z_{0}}^{\infty}\,dz\,{\cal 
K\/}_{1}(z)\,\lambda\left(s,M^{2},m_{\pi}^{2}\right)
\sigma(s).
\label{tabsrate}
\end{eqnarray}
The lower limit on the integral is $z_{0}$ = MAX[$(m_{\pi}+m_{\rho})/T 
,(m_{\pi}+M)/T$]
and the cross section 
is the same as that which was used in Eq.~(\ref{dgamma}).
Numerical results are displayed in Fig.~\ref{virtualmfp} as the
dashed curve.
The limit $M\rightarrow 0$ (and $m_{\ell}\rightarrow 0$) provides a 
consistency
check for the results.  Since the virtual photon's spin-averaged
cross section is smaller by one-third due to polarization-state
differences, the mean free path is greater by the same factor as compared
with real photon results.
Effects from the $\rho$ have been emphasized, but a straightforward
generalization would lead to structure for $\omega$ and $\phi$.

Fig.~\ref{virtualmfp} also includes the mean free path which results
from adding to the absorption rate another term which is due to thermal
reconversion to hadrons and subsequent hadronic decay.  
The thermal decay rate is
\begin{eqnarray}
\overline{\Gamma}(\gamma^{*}\rightarrow \pi^{+}\pi^{-}) & = & 
\left({4\pi\alpha\over\,f_{\rho}^{2}}\right)\,{m_{\rho}^{4}\over(M^{2}-m_{\rho}^{2})^{2}
+(m_{\rho}\Gamma_{\rho})^{2}}
\,\Gamma_{\rho}\,{{\cal K\/}_{1}(M/T)
\over{\cal K\/}_{2}(M/T)}.
\label{tdr}
\end{eqnarray}
The net mean free path is reported as the solid curve which, at the
rho peak, is about 10 fm.  This is small enough to have important 
consequences for dilepton production.

We next illustrate the consequences of reabsorption and rehadronization of
virtual photons by considering the net yield from
thermal $\pi^{+}\pi^{-} \rightarrow e^{+}e^{-}$ production
near the rho peak.  
Since the sought-after $\gamma^{*}$ propagates for 10 fm/$c$ or
longer, partly within the fireball and partly in free space, its survival 
probability could be diminished.
The net (thermal) yield from a fireball can be approximated
as the ``bare''
production rate at average temperature $\langle T \rangle$ = 170 MeV,
times an overall four-volume $V\tau$ = 
3$\times$10$^{4}$ fm$^{4}$\cite{kampfer}, times a 
survival probability ${\cal N\/}$,
\begin{eqnarray}
\left({d\/N\over\,dM}\right)^{\,net\/} & = &
\left({d\/R\over\,dM^{2}}\right)^{\,bare\/}\,\left[2M\,V\tau\right]\,{\cal N\/}.
\end{eqnarray}
This approximation is only intended to
benchmark the size of the effects and not for direct comparison
to heavy-ion data.  A dynamical model would be useful. 

The bare rate describes
dilepton production via $\pi^{+}\pi^{-}\rightarrow
\rho^{\,0}\rightarrow \gamma^{*} \rightarrow e^{+}e^{-}$\cite{wong}
\begin{eqnarray}
\left({dR\/\over\,dM^{2}}\right)^{bare} & = & 
{\sigma(M)\,M^{2}\over\,2(2\pi)^{4}}\sqrt{1-{4m_{\pi}^{2}\over\,M^{2}}}
\,T\,M\,{\cal K\/}_{1}(M/T),
\nonumber\\
\end{eqnarray}
with
\begin{eqnarray}
\sigma(M) & = & {4\pi\over 3}{\alpha^{2}\over M^{2}}\sqrt{1-{4m_{\pi}^{2}\
\over\,M^{2}}}\,\sqrt{1-{4m_{\ell}^{2}\over\,M^{2}}}\,\left(1+{2m_{\ell}^{2}
\over\,M^{2}}\right)|F_{\rho}(M)|^{2}.
\end{eqnarray}
The survival probability is approximated (using a fixed temperature
for simplicity) as
\begin{widetext}
\begin{eqnarray}
{\cal N\/} & = &
\left\lbrace
\begin{array}{ll}
\left({1\over\,t_{h}}\int_{0}^{t_{h}}\,dt\,e^{-\overline{\Gamma}^{\,tot}\,t}
\right)
\left[{e^{\Gamma^{d}t_{h}}
\over\,(t_{d}-t_{h})}\int_{t_{h}}^{t_{d}}\,dt\,e^{-\Gamma^{d}\,t}
\right]
& \quad t_{h} < t_{d} \\
\left({1\over\,t_{d}}\int_{0}^{t_{d}}\,dt\,e^{-\overline{\Gamma}^{\,tot}\,t}
\right) 
& \quad t_{d} < t_{h},
\end{array}
\right.
\label{survivalprob}
\end{eqnarray}
\end{widetext}
where $\overline{\Gamma}^{\,tot}$ is the sum of the thermal absorption and 
thermal
decay rates from Eqs.~(\ref{tabsrate}) and (\ref{tdr}), $t_{h}$ is the time 
spent by the virtual photon within the hot hadronic matter, 
$t_{d}$ is the lifetime (in the frame of the fireball) of the virtual photon 
[$t_{d}$ = $\overline\gamma\tau_{d}$, where $\tau_{d}$ is the
lifetime (plotted in Fig.~\ref{vtau})] and 
the mean Lorentz $\gamma$ factor can be written as
\begin{eqnarray}
\overline\gamma & = & {3T\over M} + 
{{\cal\,K\/}_{1}(M/T)\over{\cal\,K\/}_{2}(M/T)}.
\end{eqnarray}
Finally, $\Gamma^{d}$ is the
free-space decay rate into $\pi^{+}\pi^{-}$---the primary reason for
loss.  

\begin{figure}
\includegraphics{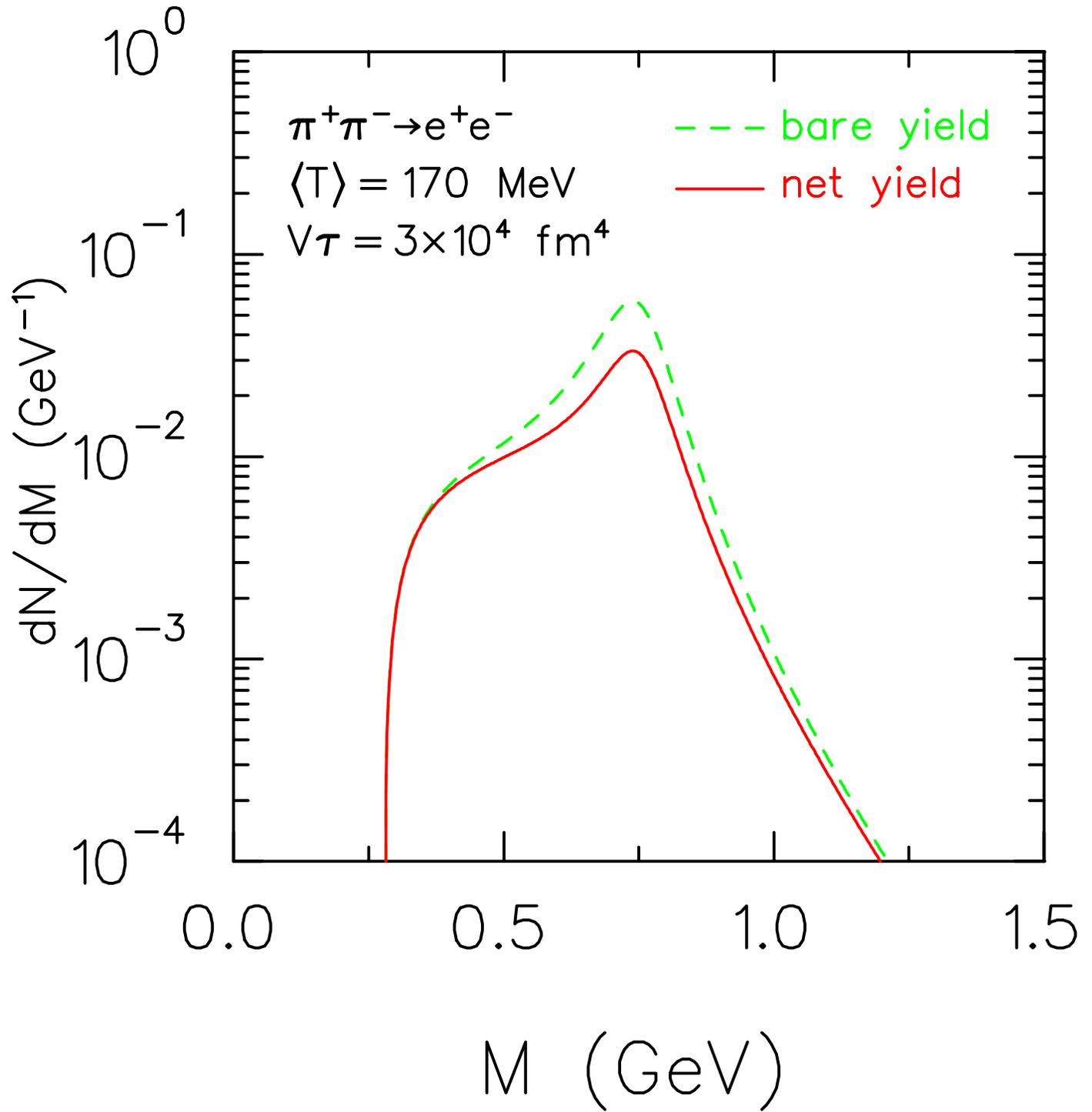}
\caption{\label{prate} Dilepton yield 
             from $\pi\pi\rightarrow\,e^{+}e^{-}$ assuming
             an average temperature $\langle${\it T\/}$\rangle$ = 170 MeV and
             a four-volume 3$\times$10$^{4\/}$ fm$^{4\/}$ .  The suppression 
             observed in the net yield is discussed
             in detail in the text---it is due to reabsorption and
             rehadronization.}
\end{figure}

We plot in Fig.~\ref{prate} the yield of thermal lepton pairs
emphasizing the rho region.   The bare-yield curve results from
setting the survival probability to unity while the
net yield includes loss effects and uses $t_{h}$ = 10 fm/$c$ in
the calculation of the survival probability.
Owing to the relative smallness of the absorption rate
as compared with the hadronization rate, the  result is not sensitive
to the value of $t_{h}$.
One can see that loss effects account for a 40\% suppression 
at the rho peak. 

Typical model calculations up to now have
used ``bare'' production alone, assuming that losses of any
kind were completely negligible.  
As models evolve and become more sophisticated,
tens of percent effects must be included.
The message here is that vector dominance helps as
it provides rich structure in the invariant mass spectra and provides
opportunity to do spectroscopy looking for collective effects, but at the 
same time it hurts in the sense that part of the rho peak is removed.
Drawing conclusions from heavy-ion dilepton production data regarding the 
extent to which the rho is melted or disappeared due to collision
broadening\cite{mycollpaper,dmscenario,gao} or other collective nuclear 
effects is unfortunately much more difficult in light  of these results.

We remark that this sort of suppression would not be seen in
muon pair spectra near the $J/\psi$ mass region since the 
$D\bar{D\/}$ threshold is too high to allow rehadronization.  

Electromagnetic probes are exceptional in terms of their ability to
penetrate and probe hot hadronic matter.  We have argued that real
photon signals suffer practically no disturbance after being produced.
Dilepton signals, which originate from intermediate virtual (off-shell)
photons, exhibit mean free paths of the order ten fm at
the rho mass, and therefore can suffer reabsorption and rehadronization 
effects.  The net dilepton yield from thermal $\pi\pi$ annihilation
in a fireball has been shown to be reduced at the rho peak with its 
apparent structure modified.  
Future models and model 
calculations for dilepton
production near vector meson mass regions would be incomplete
without accounting for possible loss.  This work provides motivation to
study the problem in greater detail, going beyond Boltzmann approximations
for particle occupation, including Bose-enhancement effects, and with 
many hadronic processes included within, say, an effective chiral 
Lagrangian description for the hadronic matter.

\acknowledgments

It is a pleasure to thank Prof. Charles Gale for useful discussions.
This work was supported by the National Science Foundation under 
grant numbers PHY-9814247 and PHY-0098760.



\end{document}